\global\def\draftcontrol{0}

   \def\versionno{ fundamental adjoint chemical }

\catcode`\@=11

\expandafter\ifx\csname draftcontrol\endcsname\relax\global\def\draftcontrol{0}
\fi

{\count255=\time\divide\count255 by 60
\xdef\hourmin{\number\count255}
\multiply\count255 by-60\advance\count255 by\time
\xdef\hourmin{\hourmin:\ifnum\count255<10 0\fi\the\count255}}
\def\draftdate{\number\month/\number\day/\number\year\ \ \ \hourmin }

\newcommand\makepapertitle{\par
  \begingroup
    \renewcommand\thefootnote{\@fnsymbol\c@footnote}%
    \def\@makefnmark{\rlap{\@textsuperscript{\normalfont\@thefnmark}}}%
    \long\def\@makefntext##1{\parindent 1em\noindent
            \hb@xt@1.8em{%
                \hss\@textsuperscript{\normalfont\@thefnmark}}##1}%
     \newpage
     \global\@topnum\z@   
     \@makepapertitle
     \thispagestyle{empty}\@thanks
  \endgroup
  \setcounter{footnote}{0}%
  \global\let\thanks\relax
  \global\let\makepapertitle\relax
  \global\let\@makepapertitle\relax
  \global\let\@thanks\@empty
  \global\let\@author\@empty
  \global\let\@date\@empty
  \global\let\@title\@empty
  \global\let\title\relax
  \global\let\author\relax
  \global\let\date\relax
  \global\let\and\relax
  \def\version{\let\version\@version\@gobble}
}
\def\@makepapertitle{%
  \newpage
   \ifnum\draftcontrol=1 {}
   \version\versionno
   \vskip 3em%
   \else
   \hfill\hbox to 3cm {\parbox{4cm}{\@pubnum}\hss}%
   \vskip 3em%
   \fi
   \begin{center}%
   \let \footnote \thanks
     {\LARGE {\@title}}%
     \vskip 1.5em%
     {\normalsize
       \lineskip .5em%
       \begin{tabular}[t]{c}%
         \@author
       \end{tabular}\par}%
     \vskip 1.5em%
     {\@bstract}%
     \end{center}%
   \par
}

\gdef\@pubnum{}
\def\pubnum#1{%
  \gdef\@pubnum{#1}}

\gdef\@bstract{}
\def\Abstract#1{%
  \gdef\@bstract{%
   \parbox{\textwidth-0pc}{%
   \centerline{\bf Abstract}\penalty1000%
\kern.2cm%
\noindent
\renewcommand\baselinestretch{1.0}%
{#1}}}
}

\def\ps@paper{\let\@mkboth\@gobbletwo%
     \ifnum\draftcontrol=1
    \def\@oddfoot{\hbox to \textwidth{\tiny \versionno \hfil\tiny\draftdate}%
    \hskip -\textwidth \hbox to \textwidth{\hfil\rm\thepage\hfil}}%
     \else\def\@oddfoot{\hbox to \textwidth{\hfil\rm\thepage\hfil}}
     \fi
     \let\@evenfoot\@oddfoot
}

\def\body{\clearpage
          \pagestyle{paper}
    }

\def\@version#1{\ifnum\draftcontrol=1
\typeout{}\typeout{#1}\typeout{}
\vskip3mm\centerline{\hbox{\fbox{\normalsize{\tt DRAFT -- #1 -- }
                   {\draftdate}}}}\vskip3mm
\fi}
\let\version\@version
\long\def\eqlabel#1{\ifnum\draftcontrol=1
                    \tag@false  
                    \tag*{(\theequation) \hbox to -0.2cm{\hspace{0cm}\small{#1}\hss}}
                    \refstepcounter{equation}
                    \edef\@currentlabel{\theequation}
                    \ltx@label{#1}          
                    \else
                    \label{#1}
                    \fi
                    }
\let\st@bibitem\@bibitem
\let\st@lbibitem\@lbibitem
\ifnum\draftcontrol=1
  \def\@bibitem#1{%
    \st@bibitem{#1}\a@@label{#1}\ignorespaces}
  \def\@lbibitem[#1]#2{%
    \st@lbibitem[#1]{#2}\a@@label{#2}\ignorespaces}
  \def\a@@label#1{%
    \gdef\a@lab{\smash{\normalfont\small#1}}
    \ifvmode
      \if@inlabel
        \global\setbox\@labels\hbox{%
          \llap{\a@lab\let\a@lab\relax
                \kern\@totalleftmargin\kern\marginparsep}%
          \box\@labels}%
      \fi
    \fi}
\fi

\documentclass[12pt,letterpaper]{article}

\usepackage{amsmath,amssymb,array,calc,rotating,epsfig,psfrag}
\usepackage[nosort]{cite}

\ifnum\draftcontrol=1
\fi

\renewcommand\section{\@startsection {section}{1}{\z@}%
                                   {-3.5ex \@plus -1ex \@minus -.2ex}%
                                   {2.3ex \@plus.2ex}%
                                   {\normalfont\large\bfseries}}
\renewcommand\subsection{\@startsection{subsection}{2}{\z@}%
                                   {-3.25ex\@plus -1ex \@minus -.2ex}%
                                   {1.5ex \@plus .2ex}%
                                   {\normalfont\normalsize\bfseries}}
\renewcommand\subsubsection{\@startsection{subsubsection}{3}{\z@}%
                                   {-3.25ex\@plus -1ex \@minus -.2ex}%
                                   {1.5ex \@plus .2ex}%
                                   {\normalfont\normalsize\it}}
\renewcommand\paragraph{\@startsection{paragraph}{4}{\z@}%
                                   {-3.25ex\@plus -1ex \@minus -.2ex}%
                                   {1.5ex \@plus .2ex}%
                                   {\normalfont\normalsize\bf}}

\numberwithin{equation}{section}

\def\ie{{\it i.e.}}

\def\revise#1       {\raisebox{-0em}{\rule{3pt}{1em}}%
                     \marginpar{\raisebox{.5em}{\vrule width3pt\
                     \vrule width0pt height 0pt depth0.5em
                     \hbox to 0cm{\hspace{0cm}{%
                     \parbox[t]{4em}{\raggedright\footnotesize{#1}}}\hss}}}}

\newcommand\nxt[1]  {\\\fnxt#1}

\def\cala         {{\cal A}}

\def\call         {{\cal L}}

\def\calo         {{\cal O}}

\def\k          {{\mathbf k}}

\def\del          {\partial}

\def\tr           {\mathop{\rm Tr}}
\def\Re           {{\rm Re\hskip0.1em}}
\def\Im           {{\rm Im\hskip0.1em}}

\def\sqr#1#2{{\vcenter{\vbox{\hrule height.#2pt
 \hbox{\vrule width.#2pt height#1pt \kern#1pt
 \vrule width.#2pt}\hrule height.#2pt}}}}

\newcommand{\ft}[2]{{\textstyle{\frac{#1}{#2}}}}

\def\dd{\delta}
\def\e{\epsilon}

\def\a{\alpha}
\def\b{\beta}

\def\hT{\hat{T}}
\def\hJ{\hat{J}}
\def\hY{\hat{Y}}
\def\l{\lambda}
\def\t{\tau}
\def\ha{\hat{a}}
\def\w{\omega}

\catcode`\@=12

\begin{document}


\title{Dynamical stabilization of runaway
potentials and landscape of vacua at finite density}

\pubnum{%
UWO-TH-07/05\\
}

\author{
Alex Buchel$^{1,2}$, Junji Jia$^{2}$, and 
V.A.Miransky$^{2,3,}$
\footnote{On 
leave from Bogolyubov Institute for 
Theoretical Physics, 03143 Kiev, Ukraine}\\[0.4cm]
\it $^1$Perimeter Institute for Theoretical Physics\\[0.4cm]
\it Waterloo, Ontario N2J 2W9, Canada\\[0.1cm]
\it $^2$Department of Applied Mathematics\\
\it University of Western Ontario\\
\it London, Ontario N6A 5B7, Canada\\[0.1cm]
\it $^3$Yukawa Institute for Theoretical Physics\\ 
\it Kyoto University, Kyoto 606-8502, Japan}

\Abstract{
We study a $SU(2)$ gauge theory with a classical complex  modulus. 
Introducing a chemical potential
for a conserved modulus hypercharge causes it to become unstable and
start condensing. We show that the
modulus condensation in turn generates
homogeneous but anisotropic
non-abelian field strength condensates. The existence of a stable
vacuum at the end point of the condensation process
depends on a modulus representation under the gauge group. For a
modulus in the fundamental representation,
the {\it global} vacuum of the theory is a state both with the rotational
symmetry and the electromagnetic $U(1)_{em}$ being spontaneously
broken. In other words, the system describes an anisotropic 
superconducting medium.
We further explore the landscape of vacua of this theory and identify  
metastable vacua with an abnormal number of Nambu-Goldstone bosons. 
The $SO(2)$ symmetry of these vacua corresponds to locking gauge,
flavor, and spin degrees of freedom. There are also metastable 
$SO(3)$ rotationally invariant vacua.
For a modulus in the adjoint representation, we show that the theory does not
have stable vacua with homogeneous anisotropic non-abelian field
strength condensates, although there are metastable vacua. The reason
of that is connected with a larger number of the physical components  
of the modulus in the case of the adjoint representation as compared
to the fundamental one.}


\makepapertitle

\body

\version\versionno

\section{Introduction}

Dynamics in
relativistic field theories with chemical potential for bosonic matter
is rich and quite sophisticated 
\cite{kh,ms,gms,Gorbar:2005pi,Brauner,Splittorff:2006uu}. 
For example, in these theories the
phenomenon of spontaneous symmetry breaking with an abnormal number of
Nambu-Goldstone (NG) bosons was revealed \cite{ms}. In
Ref. \cite{gms}, a new Higgs-like phase was described with condensates
of both gauge and scalar fields, which break gauge, flavor, and
rotational symmetries. It is noticeable that both these phenomena have
already found applications in studies of dense quark matter
\cite{abn,vect,vect1,Gorbar:2007vx}.

In this paper, we consider an essentially soluble (in the weak coupling
limit) 3+1 dimensional 
$SU(2)$ gauge model
with a (classically) complex modulus $\Phi$ in
some irreducible representation of the gauge group (no
fermions are included). 
We introduce a chemical
potential $\mu$ associated with the conserved charge of
the flavor (hypercharge)
$U(1)_Y$   symmetry. Our goal is to describe the landscape
of the ground states in this model.

The
Lagrangian density of the model is
\begin{equation}
\call=-\frac{1}{4}F_{\mu\nu}^{(a)}F^{\mu\nu(a)}+[(D_{\mu}-i\mu\delta_{\mu0})\Phi]^{\dag}[(D_{\mu}-i\mu\delta_{\mu0})\Phi]\,.
\eqlabel{lag}
\end{equation}
The gauge fields are $A_{\mu}=A_{\mu}^{(a)}T^a$, with $T^a$
being the $SU(2)$ Lie algebra generators in the representation of $\Phi$.
The field strength tensor and the covariant
derivatives are given by
\begin{equation}
F_{\mu\nu}^{(a)}=\del_\mu A_\nu^{a}-\del_\nu A_{\mu}^{a}+g \e^{abc} A_\mu^b A_\nu^c\,,\qquad D_\mu=\del_\mu-ig A_\mu\,.
\eqlabel{fs}
\end{equation}
We consider vacua of the theory \eqref{lag} with homogeneous but
generically anisotropic condensates of non-abelian field strength
$F^{(a)}_{\mu\nu}$ and with a homogeneous condensate of the complex
scalar $\Phi$. The  effective potential describing such vacua is
obtained by setting all field derivatives to zero. We find
\begin{equation}
\begin{split}
V=&-\frac{g^2}{2}\left[\left( A_0^{(a)}A_0^{(a)}\right)\left(
A_i^{(a)} A_i^{(b)}\right) -\left( A_0^{(a)}A_i^{(a)}\right)\left( A_0^{(b)}
A_i^{(b)}\right)\right]+\frac{g^2}{4}\left( A_i^{(a)}A_i^{(a)}\right)\left( A_j^{(b)}
A_j^{(b)}\right)\\
&-\frac{g^2}{4}\left( A_i^{(a)}A_j^{(a)}\right)\left( A_i^{(b)} A_j^{(b)}\right)
-\mu^2\Phi^{\dag}\Phi-2g\mu\Phi^{\dag}A_0^{(a)}T^a\Phi-\frac{g^2}{4}A_{\mu}^{(a)}A^{\mu
(a)}\Phi^{\dag}\Phi\,.
\end{split}
\eqlabel{eqvacp}
\end{equation}
Note that $-\mu^2\Phi^{\dag}\Phi$ has the form of a mass term with the
wrong sign. Because of that, the modulus $\Phi$ should start
condense. Our aim is to describe the end point in this condensing
process.

In section \ref{fundamental}, we study classical vacua of the theory with a modulus in the fundamental representation
of the gauge group. In this case,
condensing the modulus in turn leads to the generation of 
homogeneous but anisotropic
non-abelian field strength condensates. We show that the stable
vacuum of the theory is a state both with the rotational
symmetry and the electromagnetic $U(1)_{em}$ being spontaneously
broken. 
In other
words, the system describes an anisotropic superconducting medium.
We emphasize that it is the global vacuum. This result is quite
nontrivial. Indeed, there are 10 physical fields in the model,
and the problem is equivalent of studying the geometry of
a ten dimensional hypersurface corresponding to a 
{\it physical} effective potential. The latter is derived
by imposing the Gauss law constraints on potential $V$ (\ref{eqvacp})
(see Sec. \ref{fundamental} below). 
The fact that the global minimum of this hypersurface
corresponds to a complicated system where most of the initial symmetries 
are spontaneously broken is noticeable.
\footnote{This solution has been considered in our short note
\cite{bjm}. However, it was only shown there that it is a local
minimum. The question about the global vacuum
in the model remained unresolved.} 

We further explore the landscape of vacua of this theory and identify  
metastable vacua with an abnormal number of Nambu-Goldstone bosons. The
$SO(2)$ symmetry of these vacua corresponds to locking gauge, flavor,
and spin degrees of freedom. There are also metastable 
$SO(3)$ rotationally invariant vacua. 

Thus, in this simple but nontrivial model, we show that unstable
directions in nonabelian gauge theories induced by a chemical
potential for the modulus field can be stabilized by the generation
of nonabelian gauge field strength condensates. As will be discussed
in Sec. \ref{fundamental}, the Gauss law constraints play the
crucial role in this stabilization process. 

The case with the modulus in the adjoint representation is considered
in section \ref{adjoint}.
We show that, unlike the previous case, this theory does not
have stable vacua with homogeneous anisotropic non-abelian field
strength condensates, although there are metastable vacua. 
The physics underlying so dissimilar
behaviors of these two models is connected with the different
number of physical fields. While in the case
of the fundamental representation, there is one physical (Higgs) field 
connected with the modulus, there are three physical fields relating to
the modulus in the adjoint representation. As a result, the
effective potential is unbounded from below in the latter.

In section \ref{conclusion}, the main results of the paper are
summarized. In Appendices 
\ref{timelike}, \ref{disp}, and \ref{unitary}, some useful formulas
and relations are derived.

\section{$SU(2)$ gauge theory with a fundamental complex modulus at finite density}\label{fundamental}
In this section we discuss 
the $SU(2)$ gauge theory with a classical complex modulus
in the fundamental representation. Thus, we take
\begin{equation}
T^a=\frac{\t^a}{2}\,,
\eqlabel{tf}
\end{equation}
where $\t^a$ are Pauli matrices. Henceforth we will use the unitary gauge with
\begin{equation}
\Phi^T=\left(0\,,\phi\right)\,,
\eqlabel{unitaryf}
\end{equation}
where $\phi$ is real. In this gauge, the only conserved flavor charge is the
electric charge $Q_{em} = T^3 + Y/2$.

The advantage of the unitary gauge is in that all
auxiliary (gauge dependent) degrees of freedom are removed. In
particular,
in this gauge {\it the vacuum expectation values (condensates)
$A^{(a)}_\mu$ of gauge fields are well-defined
physical quantities.} 
Also, without loss of generality, by $SO(3)$
rotations we can always set
\begin{equation}
A_1^{(3)}=A_2^{(3)}=0\,,
\eqlabel{rota}
\end{equation}
\ie, the $A_i^{(3)}$ condensate is along the $z$-axis.

Although the gauge symmetry is gone in the unitary gauge,
the theory still has constraints. In fact, it is a system with
second-class constraints, similar to the theory of a free 
massive vector field $A_{\mu}$ described by the Proca Lagrangian
(for a thorough discussion of systems with second-class 
constraints, see Sec. 2.3 in book \cite{GT}).
In such theories, while the Lagrangian formalism can be used
without introducing a gauge, the physical Hamiltonian
is obtained by explicitly resolving the constraints. In 
our case, this implies that 
to obtain the physical potential $V_{phys}^{fund}$
one has to impose the Gauss law constraints on 
$V$ in \eqref{eqvacp}. 

The Gauss constraints amount to integrating out the time-like components of the
gauge potentials $A_0^{(a)}$. This could be done by using their
equations of motion. We also find to be convenient to parameterize the
gauge potential and the modulus expectation values as
\begin{equation}
A_i^{(a)}=\frac{\mu}{g}\ a_i^{(a)}\,,\qquad \phi=\frac{\mu}{g}\ \phi_0\,.
\eqlabel{dim}
\end{equation}
Further introducing
\begin{equation}
\begin{split}
&X=\left(a_1^{(1)} a_3^{(2)}-a_3^{(1)} a_1^{(2)}\right)^2\,,
\ Y=\left(a_2^{(2)} a_3^{(1)}-a_3^{(2)} a_2^{(1)}\right)^2\,,\ Z=\left(a_1^{(1)} a_2^{(2)}-a_1^{(2)} a_2^{(1)}
\right)^2\,,\\
&w_1=X+Y\,,\qquad w_2=\sum_{i,j=1}^2 \left(a_{i}^{(j)}\right)^2\,,\qquad w_3=\sum_{i=1}^2 \left(a_3^{(i)}\right)^2\,,
\end{split}
\eqlabel{xyz}
\end{equation}
after eliminating $A_0^{(a)}$ from \eqref{eqvacp},
the physical potential takes the form
\begin{equation}
\begin{split}
V_{phys}^{fund}=&\frac{\mu^4}{4g^2}\ \frac{Q\phi_0^2
(\phi_0^2+2(w_2+w_3)-8)+Q_n}{Q_d}\,,
\end{split}
\eqlabel{vpf}
\end{equation}
where
\begin{equation}
\begin{split}
Q=&\phi_0^4 (w_2+w_3)+2 \phi_0^2 \left(\left(a_3^{(3)}\right)^2 (2 w_2+w_3)+(w_2+w_3)^2\right)
+4 \left(a_3^{(3)}\right)^4 w_2\\
&+4 \left(a_3^{(3)}\right)^2 (w_1+w_2^2+w_2 w_3)+4 (w_3+w_2) (w_1+Z)\,,
\end{split}
\eqlabel{q}
\end{equation}
\begin{equation}
\begin{split}
Q_n=&\phi_0^8 \left(a_3^{(3)}\right)^2+2 \phi_0^6 \left(2 \left(a_3^{(3)}\right)^4+3 \left(a_3^{(3)}\right)^2
(w_2+w_3)+w_1+Z\right)
+4 \phi_0^4 \biggl(\left(a_3^{(3)}\right)^6\\
&+\left(a_3^{(3)}\right)^4 (5 w_2+3 w_3)
+\left(a_3^{(3)}\right)^2 (3 w_2^2+2 w_3^2+5 w_2 w_3+2 w_1+3 Z)
\\
&+2 (w_3+w_2) (w_1+Z)\biggr)
+8 \phi_0^2 \biggl(2 \left(a_3^{(3)}\right)^6 w_2+\left(a_3^{(3)}\right)^4 (4 w_2^2+3 w_2 w_3+2 w_1+Z)
\\
&+\left(a_3^{(3)}\right)^2 (w_2^3+2 w_2^2 w_3+w_2 w_3^2+(w_1+Z) (5 w_2+3 w_3))+(w_1+Z) ((w_2+w_3)^2\\
&+w_1+Z)\biggr)+16 \left(\left(a_3^{(3)}\right)^2 w_2+w_1+Z\right) \biggl(\left(a_3^{(3)}\right)^4 w_2
+\left(a_3^{(3)}\right)^2 (w_2^2+w_2 w_3\\
&+w_1)+(w_1+Z) (w_2+w_3)\biggr)\,,
\end{split}
\eqlabel{qn}
\end{equation}
\begin{equation}
\begin{split}
Q_d=&\phi_0^6+4 \phi_0^4 \left(\left(a_3^{(3)}\right)^2+w_2+w_3\right)+4 \phi_0^2 \biggl(\left(a_3^{(3)}\right)^4
+\left(a_3^{(3)}\right)^2 (3 w_2+2 w_3)\\
&+(w_2+w_3)^2+w_1+Z\biggr)+8 \biggl(\left(a_3^{(3)}\right)^4 w_2+\left(a_3^{(3)}\right)^2 (w_2^2+w_2 w_3+w_1)\\
&+(w_3+w_2) (w_1+Z)\biggr)\,.
\end{split}
\eqlabel{qd}
\end{equation}
The rest of this section is organized as follows. First, in
subsection \ref{bounded} we prove that the physical potential
\eqref{vpf} is bounded from below and possess a global minimum. In
subsection  \ref{minima} we prove that the global minima of
$V_{phys}^{fund}$ are vacua where the gauge field condensates
spontaneously break both the electromagnetic $U(1)_{em}$ and
the rotational $SO(3)_{rot}$,
\begin{equation}
U(1)_{em}\ \times\ SO(3)_{rot}\ \to\ SO(2)_{rot}\,.
\eqlabel{pattern}
\end{equation}
There are 3 massless Nambu-Goldstone (NG) modes
associated with spontaneous breakdown of the global
symmetry. Thus, these vacua describe an anisotropic superconducting medium. 
In subsection \ref{anomaly}, we discuss interesting
metastable vacua of the theory with an abnormal number of massless NG
bosons. Finally, in subsection \ref{so(3)}, metastable $SO(3)_{rot}$
invariant vacua are considered.

\subsection{$V_{phys}^{fund}$ is bounded from below and has global minima}\label{bounded}

Notice that $\{X,Y,Z,Q,Q_n,Q_d,w_1,w_2,w_3\}$ in \eqref{xyz}, \eqref{q}-\eqref{qd}, are explicitly non-negative.
Additionally, in the physical potential \eqref{vpf} the only negative contribution comes 
from a term in the numerator multiplied by $Q$. Thus,  run-away directions of the potential
\eqref{vpf} (if they were to exist) would have come only from the large values of the condensates that 
keep $\{\phi_0^2,w_2,w_3\}$ finite.  Recalling  definitions of $w_2$ and $w_3$ from \eqref{xyz},
we conclude that the only run-away direction could come from $a_3^{(3)}$ condensate, and only assuming that 
$\phi_0\ne 0$. 
However, rescaling 
\begin{equation}
a_3^{(3)}\to \l\ a_3^{(3)}\,,\qquad \l\to \infty\,,
\eqlabel{a33r}
\end{equation}
we find 
\begin{equation}
V_{phys}^{fund}\bigg|_{a_3^{(3)}\to \l\ a_3^{(3)}}\ \to \l^2\ \frac{\mu^4}{4g^2}\ \left(\phi_0^2+2w_2\right)\ 
\left(a_3^{(3)}\right)^2\,,
\eqlabel{rescale}
\end{equation}
which becomes infinitely large, since $\phi_0\ne 0$. 
Thus we conclude that 
$V_{phys}^{fund}$ given by \eqref{vpf} is in fact bounded from below.

\subsection{Global minima of $V_{phys}^{fund}$}\label{minima}
By using the inequality
\begin{equation} 
w_1\leq w_2w_3\,,
\eqlabel{inql}
\end{equation} 
we will first prove that all the
possible minima are either in the class 
\begin{equation}
(1):\ a_3^{(3)}=0\,,
\eqlabel{jcase1} 
\end{equation}
or 
\begin{equation}
(2):\ \phi_0=a_{1,2}^{(1,2)}=0\,,\qquad a_3^{(3)}\ne 0\,.
\eqlabel{jcase2}
\end{equation}
Then it will be shown  
that the global minima of potential \eqref{vpf} are in
the first class \eqref{jcase1}, with gauge field condensates 
spontaneously breaking both the $U(1)_{em}$ and the 
$SO(3)_{rot}$ rotational symmetry down to the $SO(2)_{rot}$.

The
inequality \eqref{inql} could be proved by using definitions
\eqref{xyz} and setting 
\begin{equation}
\begin{split}
&a_1^{(1)}=U\sin(\alpha)\cos(\theta_1)\,,\quad a_1^{(2)}=U\sin(\alpha)\sin(\theta_1)\,,\\
&a_2^{(1)}=U\cos(\alpha)\cos(\theta_2)\,,\quad a_2^{(2)}=U\cos(\alpha)\sin(\theta_2)\,,\\
&a_3^{(1)}=V\cos(\theta_3)\,,\quad a_3^{(2)}=V\sin(\theta_3)\,,\qquad
U,V\geq0\,.
\end{split}
\eqlabel{ju1}
\end{equation} 
Clearly, we will have
\begin{equation} 
w_1=U^2V^2\left(\sin^2\alpha\
\sin^2(\theta_3-\theta_1)+\cos^2\alpha\
\sin^2(\theta_3-\theta_2)\right)\ \le\ U^2 V^2=w_2w_3\,.
\eqlabel{ju2}
\end{equation}
To find the minima, we analyze $\frac{\partial
V_{phys}^{fund}}{\partial a_3^{(3)}}$. 
Any minimum of the potential necessarily satisfies $
\frac{\partial V_{phys}^{fund}}{\partial a_3^{(3)}}=0.$ Using
\eqref{vpf}, we have then \begin{equation}
0=a_3^{(3)}\ P\left(w_1,w_2,w_3,a_3^{(3)},\phi_0\right)\,,
\eqlabel{emoa33}
\end{equation}
where $P$ is a polynomial of $\{w_1,w_2,w_3,a_3^{(3)},\phi_0\}$:
\begin{equation}
P=32\left(w_2w_3-w_1\right)\phi_0^4\left(\phi_0^4+2\phi_0^2\left(w_2+w_3\right)+4(w_1+Z)\right)+\hat{P}\,,
\eqlabel{pdef}
\end{equation}
with polynomial $\hat{P}\equiv\hat{P}(w_1,w_2,w_3,a_3^{(3)},\phi_0)$ being manifestly non-negative.
From \eqref{inql}, we can see that $P$ is  semi-positive
definite. Moreover, we find that $P=0$ (with $a_3^{(3)}\ne 0$) implies
\begin{equation}
\phi_0=w_2=0\,.
\eqlabel{ju5}
\end{equation}
Notice that  \eqref{ju5} further implies that $w_1=Z=0$. 
Therefore the solution to \eqref{emoa33}
is indeed given either by \eqref{jcase1} or \eqref{jcase2}.

Solution \eqref{jcase2} is a trivial (unstable) vacuum with zero vacuum energy.

With \eqref{jcase1}, the physical potential
\eqref{vpf} simplifies
\begin{equation}
V_{phys}^{fund}=\frac{\mu^4}{g^2}\biggl[\frac{\phi_0^2(w_2+w_3)(\phi_0^2+2(w_2+w_3)-8)}{4(\phi_0^2+2(w_2+w_3))}
+\frac 12(w_1+Z)\biggr]\,. \eqlabel{simp}
\end{equation}
It is straightforward to analyze \eqref{simp}. It has a global
minimum
\begin{equation}
\min\ \left\{V_{phys}^{fund}\bigg|_{a_3^{(3)}=0}\right\}=-\frac{\mu^4}{2g^2}\equiv
V_{phys}^{fund\_min}\,, \eqlabel{gmin}
\end{equation}
attained with
\begin{equation}
\phi_0^2=2\,,\qquad w_2+w_3=1\,,\qquad w_1=Z=0\,.
\eqlabel{minvalues}
\end{equation}
We emphasize that for large values of the chemical potential ---
$\mu$ being much larger than the strong coupling scale of the gauge
theory --- the gauge coupling $g$ is self-consistently weak and
quantum correction to \eqref{gmin} are suppressed.
\footnote{A comment concerning the possibility of the one-loop Coleman-Weinberg (CW)
effect \cite{cw} in this model is in order.
A one-loop Coleman-Weinberg (CW)
potential would generate an effective quartic coupling of order
$g^4$. However,
the important difference between the CW model (zero
chemical potential) and the present one (finite chemical potential)
is that while there are no tree level contributions from gauge
fields in the former, the dominant effect in the present model is 
based on the tree level contributions of gauge bosons (vector condensates). 
Therefore, while in CW model the one-loop contribution of gauge fields is 
dominant, it is subleading in our case.  
In fact, one can show that a one-loop CW
effective quartic potential for a complex fundamental modulus 
will modify our classical analysis at order ${\cal O}(g^2 \ln g^2)$. This
correction is small at weak coupling.}

Let us show that in vacua \eqref{minvalues} the rotational
symmetry is broken to the $SO(2)_{rot}$ subgroup. Recalling the
definitions  \eqref{xyz}, the most general parameterization of the
vector potential condensates takes form
\begin{equation}
\begin{split}
&a_1^{(1)}=\sin\theta_1\ \cos\psi\,,\qquad a_2^{(1)}=\cos\theta_1\
\sin\theta_2\ \cos\psi\,,
\qquad a_3^{(1)}=\cos\theta_1\ \cos\theta_2\ \cos\psi\,,\\
&a_1^{(2)}=\sin\theta_1\ \sin\psi\,,\qquad a_2^{(2)}=\cos\theta_1\
\sin\theta_2\ \sin\psi\,,
\qquad a_3^{(2)}=\cos\theta_1\ \cos\theta_2\ \sin\psi\,,\\
\end{split}
\eqlabel{mincond}
\end{equation}
with arbitrary $\{\theta_1,\theta_2,\psi\}$. But from
\eqref{mincond} vector condensates $A_i^{(1)}$ and $A_i^{(2)}$ are
collinear,
\begin{equation}
A_i^{(2)}=\tan\psi\ A_i^{(1)}\,,
\eqlabel{psi}
\end{equation}
while $A_i^{(3)}=0$. Thus, the rotational symmetry is broken
to the $SO(2)_{rot}$ subgroup indeed. 

To understand better the physics in this vacua, it will be convenient
to introduce the charged fields 
$W^{(\mp)}_{\mu}=\frac{1}{\sqrt{2}}(A_{\mu}^{(1)} \pm iA_{\mu}^{(2)})$
with $Q_{em} = \mp 1$. Let us choose the direction of the collinear 
condensates $A_{i}^{(1)}$ and $A_{i}^{(2)}$ along the $z$-axis. Then, from
Eqs. (\ref{minvalues}) and (\ref{psi}), we find that the 
$W$ condensate is
\begin{equation}
W^{(\mp)}_{3} = \frac{\mu}{\sqrt{2}g} e^{\pm i\psi}.
\eqlabel{W}
\end{equation}
Therefore the electromagnetic $U(1)_{em}$ is spontaneously broken 
and these vacua are anisotropic superconducting media. 

It is noticeable that solution (\ref{W}) describes a nonzero
field strength $F_{\mu\nu}^{(a)}$ which corresponds to the presence of
{\it non-abelian} constant ``chromoelectric''-like condensates in the
ground state. In order to see this, note that, as follows from
the Gauss constraint, $A^{(3)}_0 = \mu/g$ 
for this solution (see Appendix \ref{timelike}). Then,
choosing for simplicity
the vacuum with the angle $\psi =0$,
one finds
\begin{eqnarray}
E_{3}^{(2)} &=& F_{03}^{(2)} = g\,A^{(1)}_3 A^{(3)}_0
= \frac{\mu^{2}}{g}\,.
\end{eqnarray} 
We emphasize that while an abelian constant electric field in
different media always leads to an instability,
\footnote{In metallic and superconducting
media, such an instability is classical in its origin.
In semiconductors and insulators, this instability is
manifested in creation of electron-hole
pairs through a quantum tunneling process.}
non-abelian constant chromoelectric fields do not in many cases. 
For a
discussion of the stability problem for constant non-abelian fields,
see Refs. \cite{vect} and \cite{bw}.  On a
technical side, this difference is connected with that while a vector
potential corresponding to a constant abelian electric field depends
on spatial and/or time coordinates, a constant non-abelian
chromoelectric field is expressed through constant vector potentials,
as takes place in our case, and therefore momentum and energy are good
quantum numbers in the latter.

The mass spectrum of excitations in this vacuum can be found
from Lagrangian density (\ref{lag}) 
by evaluating zeroes of the determinant of the quadratic form
of small fluctuations around the vacuum solution,
\begin{equation}
A_\mu^{(a)}(x)=\frac{\mu}{g}\ a_{\mu}^{(a)}+\dd
A_{\mu}^{(a)}\,,\qquad \Phi^T(x)=\left(0,\mu/g\ \phi_0+\dd
\phi(x)/\sqrt{2}\right) \eqlabel{fl}.
\end{equation}
One finds 10
physical states, 7 massive ones and 3 massless NG
bosons
associated with spontaneous breakdown of the global symmetries:
\begin{equation}
\begin{split}
&m^2\ =\ 0\,,\ [\times 3]\,,\\
&m^2\ =\ 2\mu^2\,,\ [\times 2]\,,\\
&m^2\ =\ 5\mu^2\,,\ [\times 2]\,,\\
&m^2\ =\ 4\mu^2\,,\ [\times 1]\,,\\
&m^2\ =\ \left(\frac 52+\frac{\sqrt{13}}{2}\right)\mu^2\,,\ [\times 1]\,,\\
&m^2\ =\ \left(\frac{5}{2}-\frac{\sqrt{13}}{2}\right)\mu^2\,,\ [\times 1]\,,\\
\end{split}
\eqlabel{spectrum}
\end{equation}
where the square bracket denotes the multiplicity of states. Notice
that in this vacuum the number of the massless NG bosons coincides with the 
number of moduli parameterizing the global minima \eqref{mincond}.

\subsection{Metastable vacuum with abnormal number of massless NG bosons}\label{anomaly}
Besides the 3 dimensional moduli space of the global minima with the
rotational $SO(3)_{rot}$ broken down to $\ SO(2)_{rot}$, 
discussed in the previous subsection, this model
has an intricate structure of 
metastable vacua, some of which were discussed in our note
\cite{bjm}.
In this subsection, we identify interesting
metastable vacua of this theory with an abnormal number of massless NG bosons.

Consider gauge field condensates such that $\vec{a}^{(3)}\equiv \{a_i^{(3)}\}=\vec{0}$, and with
vectors $\vec{a}^{(1)}\equiv \{a_i^{(1)}\}$
and $\vec{a}^{(2)}\equiv \{a_i^{(2)}\}$ satisfying\footnote{An obvious generalization would be
to consider vacua such that $\vec{a}^{(a)} \cdot \vec{a}^{(b)}=\cala^2\ \dd^{ab} $, for $\{a,b\}=\{1,2,3\}$.
However, such vacua do not exist.}
\begin{equation}
\left|\vec{a}^{(1)}\right|=\left|\vec{a}^{(2)}\right|=\cala\,,\qquad \vec{a}^{(1)} \cdot \vec{a}^{(2)}=0\,,
\eqlabel{2or}
\end{equation}
where $\cala$ is a constant. Such vacua
indeed exist:
\begin{equation}
\cala=\frac{\sqrt{2}(3-\sqrt{3})}{3}\,,\qquad \phi_0=\frac 23\ \sqrt{6(\sqrt{3}-1)}\,,
\eqlabel{sol2or}
\end{equation}
with vacuum energy necessarily higher than that of the global minimum \eqref{gmin}
\begin{equation}
V_{phys}^{fund}=-\frac{\mu^4}{g^2}\
\frac{8(2\sqrt{3}-3)}{9}\
>\ V_{phys}^{fund\_min}= -\frac{\mu^{4}}{2g^2}\,. \eqlabel{vj}
\end{equation}
We show below that vacua \eqref{2or} are perturbatively stable, thus they are metastable.

Clearly, vacua \eqref{2or} completely break the $SO(3)_{rot}$ symmetry and form a 3 parameter
family. Naively, the full pattern of the global symmetry breaking takes the form
\begin{equation}
U(1)_{em}\ \times\ SO(3)_{rot}\ \to\ 1\,,
\eqlabel{pattern1}
\end{equation}
and thus one would expect 4 massless NG bosons. In fact, there is a $SO(2)$
symmetry in the  vacua \eqref{2or} whose generator is a linear
combination
of the gauge $SU(2)$
generator $\hT_3$, the hypercharge $\hY$, and the generator $\hJ_3$ 
of the $SO(3)_{rot}$: it will be appropriate to call it a
gauge-flavor-spin locked (GFSL) symmetry.
Thus the correct pattern of the
symmetry breaking takes the form
\begin{equation}
SU(2)_{gauge}\ \times\ U(1)_Y\ \times\ SO(3)_{rot}\ \to\ SO(2)_{GFSL}\,.
\eqlabel{pattern2}
\end{equation}

To describe explicitly the $SO(2)_{GFSL}$ symmetry,
let's choose, without loss of generality, a representative vacuum of \eqref{2or} as
\begin{equation}
a_1^{(1)}=a_2^{(2)}=\cala\,.
\eqlabel{repr}
\end{equation}
Notice that the condensates $a_i^{(j)}$ with $\{i,j=1,2\}$,
\begin{equation}
\ha\equiv \left( \begin{array}{cc}
a_1^{(1)} & a_1^{(2)}  \\
a_2^{(1)} & a_2^{(2)}
 \end{array} \right)\,,
\eqlabel{mata}
\end{equation}
are closed under the action of the generators $\hT_3$
and $\hJ_3$.
In parameterization \eqref{mata}, our representative vacuum \eqref{repr} is given by
\begin{equation}
\ha_{meta}\equiv \cala\ \left( \begin{array}{cc}
1 & 0  \\
0 & 1
\end{array} \right)\,.
\eqlabel{matam}
\end{equation}
It is straightforward to compute
\begin{equation}
\begin{split}
&e^{i\a \left(\hT_3+\ft 12 \hY\right)+i\b \hJ_3}\ \ha_{meta}\, =  
U(\a)\ha_{meta}U^{T}(\b)\\ 
&=\cala\ \left( \begin{array}{cc}
\cos(\a-\b) & \sin(\a-\b)  \\
-\sin(\a-\b) & \cos(\a-\b)
\end{array} \right)\,,
\end{split}
\eqlabel{tmatam}
\end{equation}
where $\a,\b$ are arbitrary angles and
\begin{equation}
U(\a)=\left( \begin{array}{cc}
\cos\a & \sin\a  \\
-\sin\a & \cos\a
 \end{array} \right)\,.
\end{equation}
Recall that
while gauge bosons carry no hypercharge, the hypercharge
of the scalar $\Phi$ equals $+1$. Then,
\begin{equation}
e^{i\a \left(\hT_3+\ft 12 \hY\right)+i\b \hJ_3}\ \left( \begin{array}{c}
0  \\
\phi_0
\end{array} \right)\ =  \left( \begin{array}{c}
0  \\
\phi_0
\end{array} \right)\,.
\eqlabel{tphi}
\end{equation}
From Eqs. \eqref{tmatam} and \eqref{tphi} we see that the vacuum
$\{\ha_{meta},\phi_0\}$ is invariant under
transformations (\ref{tmatam}) with $\a = \b$, $i.e.$, 
under the $SO(2)$ transformations with the generator
\begin{equation}
\hat{G}_{SO(2)_{GFSL}}=\hT_3+\frac 12 \hY+\hJ_3\,,
\eqlabel{gac}
\end{equation}
which corresponds to locking gauge, flavor, and spin degrees of
freedom (compare with the color-flavor locking phase in dense
QCD \cite{Alford:1998mk}).

The mass spectrum of excitations in vacuum \eqref{matam}
(including their degeneracies) is
\begin{equation}
\begin{split}
&m^2=0,\qquad [\times 2]\,, \\
&m^2=\frac 43 \mu^2,\qquad [\times 1]\,,\\
&m^2=4 \mu^2,\qquad [\times 1]\,,\\
&m^2=\frac {16}{3} \mu^2,\qquad [\times 1]\,,\\
&m^2=\left(4-\frac {4}{\sqrt{3}}\right) \mu^2,\qquad [\times 1]\,,\\
&m^2=\left(8-\frac{8}{\sqrt{3}}+\frac{4\sqrt{6(\sqrt{3}-1)}}{3}\right) \mu^2,\qquad [\times 1]\,,\\
&m^2=\left(8-\frac{8}{\sqrt{3}}-\frac{4\sqrt{6(\sqrt{3}-1)}}{3}\right) \mu^2,\qquad [\times 1]\,,\\
&m^2=\frac 83\left(\sqrt{3}-1+{\sqrt{2\sqrt{3}-3}}\right) \mu^2,\qquad [\times 1]\,,\\
&m^2=\frac 83\left(\sqrt{3}-1-{\sqrt{2\sqrt{3}-3}}\right) \mu^2,\qquad [\times 1]\,.\\
\end{split}
\eqlabel{spef}
\end{equation}
Thus we conclude that vacuum \eqref{matam} is perturbatively stable.
It is metastable since its energy is above the global minimum of
the physical potential $V_{phys}^{fund\_min}$ \eqref{gmin}. Quite
unexpectedly, this vacuum has only 2 massless NG bosons, rather than
3 as expected from the true pattern of the symmetry breaking
\eqref{pattern1}. It is the same phenomenon as that found
in a non-gauge relativistic field model at finite density in
Ref. \cite{ms}.
As we show in Appendix \ref{disp}, only one of the
massless states in \eqref{spef} has a linear dispersion relation in
the infrared, $\w\sim k$, while the other has quadratic
dispersion relation $\w\sim k^2$. So, after all, the number of
massless excitation in vacuum \eqref{matam} is in accordance with
the Nielsen-Chadha counting rule \cite{nc}. Finally, from  Appendix
\ref{disp}, the infrared dispersion relations exhibit 
$SO(2)_{GFSL}$ symmetry  \eqref{gac}.

An interesting feature of this phase is that besides a
chromoelectric field strength condensate, there exists also a
chromomagnetic one:
\begin{equation}
H_{3}^{(3)} = F_{12}^{(3)}= gA^{(1)}_1 A^{(2)}_2 = 
\frac{\mu^2}{g} \cala^2\, .
\end{equation}

It is also interesting that a possibility of the existence of
a similar, color-spin locking, phase in dense QCD has been
recently discussed in Ref. \cite{Gorbar:2007vx}. While the question 
concerning the existence of the latter is open, 
it is quite noticeable that a phase with locking gauge and spin
degrees of freedom does exist in the present model. 

\subsection{$SO(3)_{rot}$ invariant metastable vacuum of $V_{phys}^{fund}$}
\label{so(3)}

For completeness, in this subsection, we briefly describe 
$SO(3)_{rot}$ invariant metastable vacua considered in our 
note \cite{bjm} (their 
energy density is larger than that of the GFSL vacuum).

It is easy to show that these
$SO(3)_{rot}$-invariant solutions are
\begin{equation}
A^{(3)}_0=\frac{2\mu}{g}\,,\qquad \phi_0=arbitrary\,,
\label{invarf}
\end{equation}
and no other condensates. This solution is also invariant
with respect to the electromagnetic $U(1)_{em}$.

These $SO(3)$ invariant vacua (a line of moduli) with
$V_{phys}^{fund}=0$ are either metastable or unstable. 
In order to show this,
we calculated the spectrum of excitations in this vacuum.
The spectrum includes a single massless $\phi_0$-modulus
excitation and
9 massive modes assigned to three $SO(3)_{\mbox{rot}}$ triplets
(vector modes). Their masses are $m_0 =g\phi_0$ and
$m_{\pm}=(g\phi_0\pm 2\sqrt{2}\mu)$.  While the mass $m_0$ is
connected with the neutral $A^{(3)}$ vector boson ($Q_{em} = 0$),
$m_{\pm}$ are the masses of charged $W^{(\mp)}$ vector bosons 
($Q_{em} =
\mp 1$).  For a large scalar condensate $\phi_0 > 2\sqrt{2}\mu/g$, all
the masses are positive and therefore the vacua are not unstable,
although metastable: their energy density $V_{phys}^{fund} =0$
is larger than both $V_{phys}^{fund-min}$ 
in the ground state (see Eq. (\ref{gmin}))
and the energy density in the GFSL vacuum. 

On the other hand, at the values of
$\phi_0$ less than $2\sqrt{2} \mu/g$, the mass $m_{-}$ becomes
negative and
therefore the process of a crossover of particle-antiparticle levels
takes place. The latter is a signature of the Bose-Einstein
instability: at these values of $\phi_0$, the condensate of
charged $W^{(+)}$ vector bosons is dynamically generated.

Thus, in this simple but nontrivial model, we showed that unstable
directions in nonabelian gauge theories induced by a chemical
potential for the modulus field can be stabilized by the generation
of nonabelian gauge field strength condensates. Such condensates are
homogeneous but anisotropic. 

\section{$SU(2)$ gauge theory with an adjoint complex  modulus at finite density}\label{adjoint}
In this section, we discuss the $SU(2)$ gauge theory with a classical 
complex modulus in the adjoint
representation. Thus, we take
\begin{equation}
T^{1}=\frac {1}{\sqrt{2}}\left( \begin{array}{ccc}
0&  1& 0\\
1 & 0 & 1 \\
0 & 1 & 0 \end{array} \right)\,,\
T^{2}=\frac {1}{\sqrt{2}}\left( \begin{array}{ccc}
0 & -i & 0 \\
i & 0 & -i \\
0 & i & 0 \end{array} \right)\,,\
T^{3}=\left( \begin{array}{ccc}
1 & 0 & 0 \\
0 & 0 & 0 \\
0 & 0 & -1 \end{array} \right)\,,
\eqlabel{ta}
\end{equation}
for the $SU(2)$ generators in adjoint (vector) representation,
canonically normalized as
\begin{equation}
\tr\left(T^{a}T^{b}\right)=2\ \dd^{ab}\,,\qquad \left[T^a,T^b\right]=i\ \e^{abc}\ T^c\,.
\eqlabel{lienorm}
\end{equation}
Henceforth we will use the unitary gauge with
\begin{equation}
\Phi^T=\frac{\mu}{g}\ \left(\phi_{1}+i\phi_0,0,\phi_{3}+i\phi_0\right)\,,
\eqlabel{uniraty}
\end{equation}
where $\{\phi_0,\phi_1,\phi_3\}$ are all real\footnote{One can
verify that for the vacua of interest \eqref{uniraty} provides a
good gauge choice, see Appendix \ref{unitary}.}. Notice that our
unitary gauge removes three degrees of freedom from the complex
scalar, which is expected given that the gauge group is completely
broken. As in section \ref{fundamental}, by $SO(3)$ rotations we can
always set \eqref{rota}. Again, the physical potential
$V_{phys}^{adj}$ is obtained by integrating out the time-like
components  $a^{(a)}_0$ of the gauge potentials.

The rest of the section is organized as follows. First, in
subsection \ref{runaway} we explicitly exhibit runaway directions of
the physical potential $V_{phys}^{adj}$. Thus, unlike the gauge
theory with fundamental matter, here, the runaway directions can not
be stabilized by homogeneous (though generically anisotropic)
gauge field strength condensates. The reason for that is
discussed in the next subsection.
Necessarily, in this theory any
vacuum is at most metastable. In subsection \ref{metastable}, we
identify metastable $SO(3)_{rot}$ invariant
vacua with zero energy density. In subsection (\ref{unstable}), 
we identify $SO(2)_{rot}$
invariant vacua with negative $V_{phys}^{adj}$. Unfortunately, these
vacua are perturbatively unstable.

\subsection{Runaway directions of $V_{phys}^{adj}$}\label{runaway}
We will not present here the general expression for
$V_{phys}^{adj}$. We find though
\begin{equation}
V_{phys}^{adj}\bigg|_{a_{1,2}^{(1,2,3)}=a_3^{(1,2)}=0}
=\frac{\mu^4}{g^2}\ \left[\ \left(\left(a_3^{(3)}\right)^2-1\right)
\left(\phi_1^2+\phi_3^2+2
\phi_0^2\right)+\frac{(\phi_1^2-\phi_3^2)^2}{\phi_1^2+\phi_3^2+2
\phi_0^2} \ \right]\, \eqlabel{unbounded}
\end{equation}
for arbitrary $\{a_3^{(3)},\phi_0,\phi_1,\phi_3\}$.
The runaway directions are explicit in \eqref{unbounded}.
For example,
\begin{equation}
\begin{split}
\phi_1^2+\phi_3^2<{\rm const}\,,\qquad
\left|a_{3}^{(3)}\right|<1\,,\qquad \phi_0\to \infty\,,\\
(\phi_1^2-\phi_3^2)^2 <{\rm const}\,,\qquad
\left|a_{3}^{(3)}\right|<1\,,\qquad \phi_1\to \infty\,.
\end{split}
\eqlabel{rund}
\end{equation}
In fact, at $|a_{3}^{(3)}| < 1$, any direction 
with the second term in the square brackets in 
(\ref{unbounded})
being bounded, and at least one of the fields $\phi_0, \phi_1$,
$\phi_3$ going to infinity, is a runaway one.

So, indeed the potential  $V_{phys}^{adj}$ is unbounded from below
for homogeneous (generically anisotropic) adjoint complex scalar
and gauge field strength  condensates. It seems that the
reason of this feature is connected with a larger number (three) of
physical fields connected with the modulus in this case as
compared to the fundamental one. For example, if one
removes the field $\phi_0$ (coming from the imaginary part of
the modulus) and one of the fields $\phi_1, \phi_3$  
in Eq. (\ref{unbounded}), this expression becomes bounded
from below.

\subsection{$SO(3)_{rot}$ invariant metastable vacuum of $V_{phys}^{adj}$}\label{metastable}
There are simple $SO(3)_{rot}$ invariant vacua of
$V_{phys}^{adj}$ with zero energy density:
\begin{equation}
a_{1,2,3}^{(1,2,3)}=0\,,\qquad \phi_1=\phi_0=0\,,\qquad \phi_3={\rm arbitrary}\,.
\eqlabel{azero}
\end{equation}
These vacua are also invariant under the electromagnetic $U(1)_{em}$
with the generator $Q_{em} = T^3 + Y$.

The mass spectrum of fluctuations, including degeneracies,  in vacua  \eqref{azero} is
\begin{equation}
\begin{split}
&m^2=0\,,\qquad [\times 1]\,,\\
&m^2=4 \mu^2\,,\qquad [\times 2]\,,\\
&m^2=2 \mu^2\ \phi_3^2\,,\qquad [\times 3]\,,\\
&m^2=\mu^2\ \left(\phi_3+1\right)^2\,,\qquad [\times 3]\,,\\
&m^2=\mu^2\ \left(\phi_3-1\right)^2\,,\qquad [\times 3]\,.\\
\end{split}
\eqlabel{azerofl}
\end{equation}
For generic $\phi_3$, there is a single massless $\phi_3$-modulus
excitation.
Vacua \eqref{azero}
are perturbatively stable for $\phi_3>1$. They are metastable since the physical potential
 $V_{phys}^{adj}$ is unbounded from below. When $\phi_3<1$ the mass $m_-\equiv\mu (\phi_3-1)$
is negative and therefore the process of a crossover of particle-antiparticle levels takes place,
making vacuum \eqref{azero} unstable.
The same phenomenon occurs in $SO(3)_{rot}$ invariant vacua of the $SU(2)$ gauge theory with a complex
fundamental modulus considered in subsection \ref{so(3)} above.

\subsection{$SO(2)_{rot}$ invariant unstable vacuum of $V_{phys}^{adj}$}\label{unstable}

In this subsection we consider a one parameter family of the vacua
of  $V_{phys}^{adj}$ which have the following pattern of the
spontaneous symmetry breaking
\begin{equation}
U(1)_{Y}\  \times\ SO(3)_{rot}\ \to\ SO(2)_{rot}\,,
\eqlabel{s02a}
\end{equation}  
which is similar to that considered in subsection \ref{minima}.

These vacua are characterized by the condensates:
\begin{equation}
\begin{split}
&a_{1,2}^{(1,2,3)}=a_3^{(3)}=0\,,\\
&a_3^{(1)}=-\frac{\sqrt{5\sqrt{17}-19}}{2}\ \cos\alpha\,,\qquad a_3^{(2)}=-\frac{\sqrt{5\sqrt{17}-19}}{2}\ \sin\alpha\,,
\end{split}
\eqlabel{so2cond1}
\end{equation}
\begin{equation}
\begin{split}
&\phi_0=\frac{ \sqrt{2}\ (7 \sqrt{17}-33)\ \sqrt{\sqrt{17}-3}}{4\ (1+\sqrt{17})\ \sqrt{\cos^2\a\ (7 \sqrt{17}-33)
+9+\sqrt{17}}}\ \cos\a\ \sin\a\,,\\
&\phi_1=  \frac{\sqrt{2}\ (28-4 \sqrt{17}-\sqrt{\sqrt{17}-3}\ (\cos^2\a\ (7 \sqrt{17}-33)+9+\sqrt{17}))}{4\ (1
+\sqrt{17})\ \sqrt{\cos^2\a\ (7 \sqrt{17}-33)+9+\sqrt{17}}}\,,\\
&\phi_3= \frac{\sqrt{2}\ (28-4 \sqrt{17}+\sqrt{\sqrt{17}-3}\ (\cos^2\a\ (7 \sqrt{17}-33)
+9+\sqrt{17}))}{4\ (1+\sqrt{17})\ \sqrt{\cos^2\a (7 \sqrt{17}-33)+9+\sqrt{17}}}\,,
\end{split}
\eqlabel{so2cond2}
\end{equation}
where $\a$ is an arbitrary parameter. The vacuum energy corresponding to \eqref{so2cond1}, \eqref{so2cond2} is
\begin{equation}
V_{phys}^{adj}=\frac{\mu^4}{g^2}\
\frac{17\sqrt{17}-71}{16}\ <\ 0\,. \eqlabel{vaq}
\end{equation}
Note also that because the time component $a^{(3)}_0$, determined
from the Gauss law constraint, is
nonzero for this solution, $a^{(3)}_0 = \frac 12 \sqrt{\sqrt{17}-3} $
(see Appendix \ref{timelike}), 
there are chromomagnetic field strengths condensates
$F_{03}^{(1)}$ and/or $F_{03}^{(2)}$ in these vacua. 

The mass spectrum of fluctuations in vacua \eqref{so2cond1},   \eqref{so2cond2} is
\begin{equation}
\begin{split}
&m^2=0\,,\qquad [\times 3],\\
&m^2=(5-\sqrt{17}) \mu^2\,,\qquad [\times 2],\\
&m^2=\left(\frac 72- \frac{\sqrt{17}}{2}\right)\mu^2\,,\qquad [\times 2],\\
&m^2=\left(\frac {15}{4}- \frac{\sqrt{17}}{4}\right)\mu^2\ \,,\qquad [\times 1]\,,
\end{split}
\eqlabel{azerof2}
\end{equation}
and the remaining 4 states have mass
\begin{equation}
m^2=\lambda\ \mu^2\,,
\eqlabel{m4}
\end{equation}
where $\lambda$ is a root of the following forth order polynomial
\begin{equation}
4\lambda^4+(1-7 \sqrt{17}) \lambda^3+(58 \sqrt{17}-182) \lambda^2+(993 \sqrt{17}-4151) \lambda-8382 \sqrt{17}+34578=0\,.
\eqlabel{4pol}
\end{equation}
Notice from \eqref{azerof2} that 
there are 3 massless NG bosons, as expected from the
pattern of the spontaneous symmetry breaking \eqref{s02a}.
Also, 2 of the roots of \eqref{4pol} are imaginary.
Indeed, solving numerically \eqref{4pol} we find
\begin{equation}
\lambda=\{0.5457\,; 4.397\,; 1.011\pm0.9307\ i\}\,.
\eqlabel{ns4pol}
\end{equation}
We conclude that vacua \eqref{so2cond1}, \eqref{so2cond2}, albeit  having lower vacuum energy than the
metastable vacua \eqref{azero}, are unstable.

\section{Conclusion}
\label{conclusion}

It is rather surprising that dynamics of a relatively simple model, such
as the present one, leads to such a rich landscape of 
stable and metastable vacua. In particular,
it is noticeable that the anisotropic superconducting vacuum and
the gauge-flavor-spin locking one discussed in Sec. \ref{fundamental}
have their analogues in dense quark matter
(see Refs. \cite{vect,vect1,Gorbar:2007vx,Alford:1998mk}), where
their studies are limited by complexities in infrared dynamics
in QCD. One can say that the present consideration yields a
proof that the dynamics with vector condensates of gauge fields is
a real thing.

The model with the field $\Phi$ in the fundamental representation
of the gauge group considered here is a very special case of that
in Ref. \cite{gms}, with the quartic coupling constant
$\lambda$
and the mass of the scalar field $\Phi$ chosen to be zero.
This special limit has two important advantages. First, it yields
the dynamics with moduli, and this led us to revealing the
phenomenon of the dynamical stabilization of runaway potentials
at finite density. Secondly, this limit, retaining richness of
the dynamics, simplifies the analysis of the structure
of the vacuum manifold. This allowed us to establish 
the structure of the global
vacuum and find some interesting metastable ones in the model.

We also found that
the stability properties of the gauge theory with classical complex
modulus and chemical potential are intrinsically different for
theories in the fundamental and adjoint representations: while the one
in the fundamental representation is stabilized by a condensation
of gauge fields, the theory in the adjoint one is always
unstable. The reason for that is connected
with different numbers of
physical components in the moduli in these two cases, being larger in the 
latter. A special role of the fundamental representation could
be welcome, taking into account that, after all, the electroweak
Higgs field is assigned to this representation. 

We would not be surprised if the dynamics revealed here find
applications in such different areas as cosmology and condensed
matter.

\section*{Acknowledgments}
A.B. would like to thank Maxim Pospelov for valuable discussions.
A.B. would like to thank the Aspen Center for Physics for hospitality where part  of this work was done.
A.B. research at Perimeter Institute is supported in part by the
Government of Canada through NSERC and by the Province of Ontario through MEDT.
A.B. gratefully acknowledges further support by an NSERC Discovery grant.
The
work of J.J. and V.A.M. was supported by the Natural Sciences and Engineering
Research Council of Canada. 
V.A.M. is grateful to Prof. Taichiro Kugo and Prof. Teiji Kunihiro 
for their warm hospitality during his stay
at Yukawa Institute for Theoretical Physics, Kyoto University.

\appendix

\section{Gauss Law constraints}
\label{timelike}
In this section we collect expressions for the expectation values of the time
components of non-abelian gauge potential condensates constrained by the Gauss Law.  
\nxt For the model with a complex modulus in a fundamental representation, the values of the 
gauge potential time
components in global minimum \eqref{gmin} are
\begin{equation} A_0^{(1)}=0\,, \qquad A_0^{(2)}=0\,, \qquad
A_0^{(3)}=\frac{\mu}{g}\,. \eqlabel{tcfgmin} 
\end{equation} 
\nxt For the model with a complex modulus in a fundamental representation, the values of the gauge potential time
components in metastable vacua in subsection \ref{anomaly} are
\begin{equation}  A_0^{(1)}=0\,, \qquad A_0^{(2)}=0\,, \qquad
A_0^{(3)}=\frac{2\sqrt{3}}{3}\frac{\mu}{g}\,. \eqlabel{tcfanomal}
\end{equation} 
\nxt For the model with a complex modulus in a fundamental representation, the values of the gauge potential time
components in the $SO(3)_{rot}$ invariant vacua in subsection \ref{so(3)}
are
\begin{equation}  A_0^{(1)}=0\,, \qquad A_0^{(2)}=0\,, \qquad
A_0^{(3)}=\frac{2\mu}{g}\,. \eqlabel{tcfso3} \end{equation}
\nxt For the model with a complex modulus in an adjoint representation, the values of the gauge potential time
components in $SO(3)_{rot}$
invariant vacua in subsection \ref{metastable} are
\begin{equation}
A_0^{(1)}=0\,, \qquad A_0^{(2)}=0\,, \qquad A_0^{(3)}=\frac{\mu}{g}\,.
\eqlabel{tcaso3} \end{equation} 
\nxt For the model with a complex modulus in an adjoint representation, the values of the gauge potential time
components in the $SO(2)_{rot}$ invariant vacua in subsection \ref{unstable} are
\begin{equation}
A_0^{(1)}=0\,, \qquad A_0^{(2)}=0\,, \qquad
A_0^{(3)}=\frac{\sqrt{\sqrt{17}-3}}{2}\frac{\mu}{g}\,.
\eqlabel{tcaso2}
\end{equation}

\section{Infrared dispersion relation for fluctuations \eqref{spef}}
\label{disp}

Introduce
\begin{equation}
\k_\perp^2\equiv k_1^2+k_2^2\,,\qquad \k^2\equiv\k_\perp^2+k_3^2\,.
\eqlabel{kp}
\end{equation}
Dispersion relations for the fluctuations \eqref{spef} in the infrared region $\k^2\ll \mu^2$ are
\begin{equation}
\begin{split}
\w^2=&\frac 13\ \k^2+\calo\left(\k^4\right)\,,
\end{split}
\eqlabel{st1}
\end{equation}
\begin{equation}
\begin{split}
\w^2=&\frac{7-4 \sqrt{3}}{16}\ \k^2\ \left(k_3^2+(2 \sqrt{3}+3) \k_\perp^2\right)+\calo\left(\k^6\right)\,,
\end{split}
\eqlabel{st2}
\end{equation}\begin{equation}
\begin{split}
\w^2=&\frac 43\ \mu^2+\left(\frac 53-\frac{\sqrt{3}}{3}\right) k_3^2+\left(\frac{9707}{6623}
-\frac{916\sqrt{3}}{6623}\right) \k_\perp^2+\calo\left(\k^4\right)\,,
\end{split}
\eqlabel{st3}
\end{equation}\begin{equation}
\begin{split}
\w^2=&4\ \mu^2-\left(1+\sqrt{3}\right) k_3^2+\left(\frac{71\sqrt{3}}{9}+\frac{118}{9}\right) \k_\perp^2
+\calo\left(\k^4\right)\,,
\end{split}
\eqlabel{st4}
\end{equation}\begin{equation}
\begin{split}
\w^2=&\frac{16}{3}\ \mu^2+\left(\frac{4\sqrt{3}}{3}+\frac{10}{3}\right) k_3^2+\left(6 \sqrt{3}-8\right)
\k_\perp^2+\calo\left(\k^4\right)\,,
\end{split}
\eqlabel{st5}
\end{equation}\begin{equation}
\begin{split}
\w^2=&\left(4-\frac{4}{\sqrt{3}}\right)\ \mu^2+\left(\frac 23+\frac{\sqrt{3}}{3}\right) k_3^2
+\left(11-6 \sqrt{3}\right) \k_\perp^2+\calo\left(\k^4\right)\,,
\end{split}
\eqlabel{st6}
\end{equation}
\begin{equation}
\begin{split}
\w^2=&\left(8-\frac{8}{\sqrt{3}}+\frac{4 \sqrt{6 (\sqrt{3}-1)}}{3}\right)\ \mu^2
+\left(\frac 32-\frac{\sqrt{3}}{6}+\left(\frac{\sqrt{3}}{9}+\frac 16\right) \sqrt{6 (\sqrt{3}-1)}\right) k_3^2\\
&+\left(\frac{6763}{3222}-\frac{469\sqrt{3}}{3222}+\left(\frac{617}{3222}+\frac{3583\sqrt{3}}{19332}\right)
\sqrt{6 (\sqrt{3}-1)}\right) \k_\perp^2+\calo\left(\k^4\right)\,,
\end{split}
\eqlabel{st7}
\end{equation}
\begin{equation}
\begin{split}
\w^2=&\left(8-\frac{8}{\sqrt{3}}-\frac{4 \sqrt{6 (\sqrt{3}-1)}}{3}\right)\ \mu^2
+\left(\frac 32-\frac{\sqrt{3}}{6}-\left(\frac{\sqrt{3}}{9}+\frac 16\right) \sqrt{6 (\sqrt{3}-1)}\right) k_3^2\\
&+\left(\frac{6763}{3222}-\frac{469\sqrt{3}}{3222}-\left(\frac{617}{3222}+\frac{3583\sqrt{3}}{19332}\right)
\sqrt{6 (\sqrt{3}-1)}\right) \k_\perp^2+\calo\left(\k^4\right)\,,
\end{split}
\eqlabel{st8}
\end{equation}
\begin{equation}
\begin{split}
\w^2=&\frac 83 \left(\sqrt{3}-1+\sqrt{2 \sqrt{3}-3}\right)\ \mu^2
+\left(1+\left(1+\frac{2 \sqrt{3}}{3}\right) \sqrt{2 \sqrt{3}-3}\right) k_3^2\\
&-\left(\frac{224}{37}+\frac{138\sqrt{3}}{37}+\left(\frac{190\sqrt{3}}{37}+\frac{332}{37}\right)
\sqrt{2 \sqrt{3}-3}\right) \k_\perp^2+\calo\left(\k^4\right)\,,
\end{split}
\eqlabel{st9}
\end{equation}
\begin{equation}
\begin{split}
\w^2=&\frac 83 \left(\sqrt{3}-1-\sqrt{2 \sqrt{3}-3}\right)\ \mu^2
+\left(1-\left(1+\frac{2 \sqrt{3}}{3}\right) \sqrt{2 \sqrt{3}-3}\right) k_3^2\\
&-\left(\frac{224}{37}+\frac{138\sqrt{3}}{37}-\left(\frac{190\sqrt{3}}{37}+\frac{332}{37}\right)
\sqrt{2 \sqrt{3}-3}\right) \k_\perp^2+\calo\left(\k^4\right)\,.
\end{split}
\eqlabel{st10}
\end{equation}

\section{Unitary gauge}
\label{unitary}
We show here that the unitary vacuum choice \eqref{uniraty}  is the most general.
Indeed, in the conventional (purely imaginary) choice for the $SU(2)$ generators in adjoint representation
\begin{equation}
T_c^{1}=-i\left( \begin{array}{ccc}
0&  0& 0\\
0 & 0 & 1 \\
0 & -1 & 0 \end{array} \right)\,,\
T_c^{2}=-i\left( \begin{array}{ccc}
0 & 0 & -1 \\
0 & 0 & 0 \\
1 & 0 & 0 \end{array} \right)\,,\
T_c^{3}=-i \left( \begin{array}{ccc}
0 & 1 & 0 \\
-1 & 0 & 0 \\
0 & 0 & 0 \end{array} \right)\,,
\eqlabel{tac}
\end{equation}
vacuum \eqref{uniraty} is represented by
\begin{equation}
\Phi_{c,0}=\Re\biggl(\Phi_{c,0}\biggr)+i\ \Im\biggl(\Phi_{c,0}\biggr)
=\left[ \begin {array}{c} 0\\\noalign{\medskip}-\frac{1}{\sqrt{2}}\, \left(
{ \phi_1}+{ \phi_3} \right) \\\noalign{\medskip}0\end {array}
 \right] +i\
\left[ \begin {array}{c} \frac{1}{\sqrt{2}}\, \left( { \phi_1}-{ \phi_3}
 \right) \\\noalign{\medskip}-\sqrt {2}{ \phi_0}\\\noalign{\medskip}0
\end {array} \right] \,.
\eqlabel{arb}
\end{equation}
Notice that  the  most general vacuum in canonical representation
\eqref{tac} can be parameterized by two real
vectors $\{\Re(\Phi_c),\Im(\Phi_c)\}$ in three dimensional space 
with gauge transformations acting as rotations
separately on $\{\Re(\Phi_c)\}$ and $\{\Im(\Phi_c)\}$.
Thus, to demonstrate the validity of \eqref{uniraty},
all we  need to do is to show that in the canonical representation
an arbitrary vacuum can be parameterized as in  \eqref{arb}.

But it is easy to show this. Let $\vec{v}_1\equiv \Re(\Phi_c)$ and $\vec{v}_2\equiv \Im(\Phi_c)$
are two vectors representing the generic vacuum.
First, without any loss of generality we can choose coordinate axes  in
such a way that the vector $\vec{v}_1$ is aligned along the $y$-axis.
Then, obviously we can identify $\vec{v}_1$ with $\Re(\Phi_{c,0})$.
At this stage, the vector $\vec{v}_2$ is arbitrary, but with our choice
of coordinate axes we can still do rotations about $y$-axis.
Any rotation of this type does not change 
$\vec{v}_1 =\Re(\Phi_{c,0})$. 
Clearly, no matter what $\vec{v}_2$ is,
by a $y$-rotation we can make it to have $z$-component to vanish ---
but then the rotated $\vec{v}_2$ would be identified with  $\Im(\Phi_{c,0})$.


\end{document}